\documentclass{article}

\usepackage[nonatbib, final]{neurips_2025}

\usepackage[utf8]{inputenc} 
\usepackage[T1]{fontenc}    
\usepackage{hyperref}       
\usepackage{url}            
\usepackage{booktabs}       
\usepackage{amsfonts}       
\usepackage{nicefrac}       
\usepackage{microtype}      
\usepackage{xcolor}         

\usepackage[numbers]{natbib}

\usepackage{styles/macros}

\usepackage{amsmath,amsfonts,bm}

\def\eqref#1{equation~\ref{#1}}

\def\1{\bm{1}}

\DeclareMathAlphabet{\mathsfit}{\encodingdefault}{\sfdefault}{m}{sl}
\SetMathAlphabet{\mathsfit}{bold}{\encodingdefault}{\sfdefault}{bx}{n}

\usepackage{amsmath}
\usepackage{amssymb}
\usepackage{mathtools}
\usepackage{amsthm}

\usepackage{tcolorbox}
\usepackage{caption}
\usepackage{subcaption}

\usepackage{listings}
\usepackage[compatibility=false]{caption}
\usepackage{booktabs}
\definecolor{lightblue}{RGB}{95, 158, 160} 

\usepackage[normalem]{ulem}
\usepackage{paralist}
\usepackage{mathtools}
\usepackage{xspace}
\usepackage{tikz}
\usepackage{amsmath}
\usetikzlibrary{shapes,arrows}
\usetikzlibrary{positioning}
\usetikzlibrary{arrows.meta}
\usepackage{pgfplots}
\usepackage{subcaption}
\usepackage{booktabs}
\usepackage{tcolorbox}
\usepackage{wrapfig}
\usepackage{hyperref}
\usepackage{tabularx}
\usepackage{booktabs}
\usepackage{array}

\usepackage[inline]{enumitem}
\usepackage[ruled,vlined]{algorithm2e}
\usepackage{adjustbox}

\usepackage[capitalize,noabbrev]{cleveref}

\title{Adversarial Query Synthesis via Bayesian Optimization}

\author{
    Jeffrey Tao\thanks{Equal contribution. Correspondence to: \href{mailto:jefftao@seas.upenn.edu}{jefftao@seas.upenn.edu}} 
    \quad\; Yimeng Zeng\footnotemark[1] 
    \quad\; Haydn T. Jones 
    \quad\; Natalie Maus \\
    \textbf{Osbert Bastani} \quad\; \textbf{Jacob R. Gardner} \quad\; \textbf{Ryan Marcus} \\
    University of Pennsylvania
}

\begin{document}

\maketitle

\begin{abstract}
    Benchmark workloads are extremely important to the database management research community, especially as more machine learning components are integrated into database systems. Here, we propose a Bayesian optimization technique to automatically search for difficult benchmark queries, significantly reducing the amount of manual effort usually required. In preliminary experiments, we show that our approach can generate queries with more than double the optimization headroom compared to existing benchmarks. 
\end{abstract}

\section{Introduction}
Database researchers strive to build high-performance systems for processing SQL queries. To evaluate their work, the database community depends on benchmarks to approximate challenging queries. Traditionally, synthetic benchmarks (i.e., data drawn from random distributions) like the TPC family~\cite{tpcds,tpch_analyzed} have been dominant, but some benchmarks over ``real world'' data, like the Join Order Benchmark (JOB) over the Internet Movie database (IMDb) have also been used~\cite{howgood}. Work applying machine learning techniques to database systems (e.g.,~\cite{ml_index}) has recently underscored the importance of ``real world`` benchmarks, as synthetic data distributions are normally either overly simplistic (e.g., a uniform key column) or entirely unlearnable (e.g., noise)~\cite{benchmark_lis}. As a concrete example, consider a recent report from AWS Redshift, which highlights the relative ``easiness'' of traditional benchmarks compared to the workloads they observe~\cite{redshift_workload}.

This gap has led to a scramble (or perhaps an arms race) for larger and more challenging benchmarks, with solutions including ``matching'' synthetic workloads to real traces~\cite{redbench}, manually constructing new querysets over public data~\cite{bao}, synthesizing queries with LLMs~\cite{sqlstorm}, or adding additional queries to existing datasets~\cite{job_complex}. With the exception of~\cite{sqlstorm}, all of these approaches require significant manual effort. And, with the exception of~\cite{job_complex}, all of these approaches do not optimize for the \emph{difficulty} of the benchmark (i.e., the potential room for improvement over existing systems, which we call \emph{headroom}).


In this work, we present a preliminary system that automatically synthesizes simple queries with significant optimization headroom (i.e., high difficulty). Our approach uses Bayesian optimization to discover query-and-plan pairs for which the discovered plan performs significantly better than the baseline system. In addition to being fully automated, this approach has three key advantages: first, by directly optimizing for query headroom, we show that we can find queries that have \emph{more than double} the headroom of existing benchmarks. Second, by searching over plans in addition to queries, we generate \emph{witness plans} with significantly better performance than the baseline system, which is potentially useful for debugging performance bugs. Third, by restricting our search space to simple queries (conjunctive queries with no aliased joins), our benchmark queries can run on every relational database system currently known to the authors.

At a high level, our approach formulates the benchmark building problem as a optimization problem, and uses Bayesian optimization enhanced with a latent-space representation of both SQL queries and plans. To handle the structured, discrete nature of SQL queries, we use a composite variational autoencoder that encodes queries and plans into a shared continuous latent space, enabling efficient search. Queries are embedded with a large text embedding model and decoded with a fine-tuned compact LLM using grammar-constrained decoding~\cite{xgrammar}, ensuring syntactic validity, while plans benefit from a closed vocabulary representation that guarantees well-formedness. This combination allows the system to efficiently propose and evaluate candidate pairs, iteratively refining its model to uncover challenging, high-headroom queries.
\section{Methods}
\label{sec:methods}

The core problem we address is the automated discovery of adversarial pairs $(Q, P)$, where $Q$ is a SQL query and $P$ is an execution plan for $Q$. An adversarial pair is one where the latency of plan $P$, denoted $L(P)$, is significantly lower than the latency of the plan generated by the database's default query optimizer, $L(P_{\text{default}})$. We formulate this as a black-box optimization problem aimed at maximizing either the \textbf{relative speedup}, $L(P_{\text{default}}) / L(P)$, or the \textbf{absolute speedup}, $L(P_{\text{default}}) - L(P)$.

\textbf{Black-box and Bayesian optimization.} In black-box optimization, we aim to optimize an \textit{oracle} objective function $f(\mathbf{x})$ over a space of candidates $\mathbf{x}^*=\text{argmax}_{\mathbf{x}\in\mathcal{X}} f(\mathbf{x})$. Examples of such problems include molecule activity maximization for drug discovery \citep{design-bench, lolbo}, and binding affinity of DNA sequences or proteins \citep{TFbind,lambo2}.  Commonly, $f(x)$ is assumed to be expensive to evaluate or even completely unknown.

Bayesian optimization is a sample-efficient framework to solve these expensive model-based optimization problems \citep{osborne2009, mockus1982, snoek2012}. At iteration $t$ of BO, one has access to observations $\mathcal{D}_t = \{(\mathbf{x}_{i}, y_{i})\}_{i=1}^{t}$, where $y_{i}$ denotes the objective value of the input $x_{i}$. Typically, a Gaussian process \citep{rasmussen2003gaussian} is employed as the surrogate model to approximate the objective function using these inputs and values. This surrogate model aids the optimization by employing an acquisition function, which strategically proposes the next candidates for evaluation. After querying these candidates through the true oracle, the surrogate model is updated with the new observations. This process gradually builds a more comprehensive dataset and refines the surrogate model, thereby improving the quality of the proposed samples in future iterations.

\textbf{Bayesian optimization over SQL queries.} Bayesian optimization is a sample-efficient technique for optimizing expensive black-box functions. However, standard BO operates on continuous vector spaces, whereas our search space consists of discrete, structured objects (SQL queries and query plans). We bridge this gap using Latent Space BO (LS-BO), a technique proven effective in domains like molecule design \cite{lolbo} and offline query planning \cite{bayesqo}.

The LS-BO framework requires two key components: (1) an encoder $\mathcal{E}$ that maps structured inputs into a continuous latent space, $\mathcal{Z}$, and (2) a decoder $\mathcal{D}$ that maps points from $\mathcal{Z}$ back into the structured input space. In this paper, we introduce a novel, composite Variational Autoencoder (VAE) architecture that creates a joint latent space for both queries and plans together.

We construct a continuous latent space $\mathcal{Z}=[z_q;z_p]\in\mathbb{R}^{320}$ that concatenates query VAE latents ($z_q\in\mathbb{R}^{256}$) with plan VAE latents ($z_p\in\mathbb{R}^{64}$):
\begin{align*}
\mathcal{E}_q&:\text{SQL string}\rightarrow z_q,& \mathcal{D}_q&:z_q\rightarrow \text{SQL string},\\
\mathcal{E}_p&:\text{plan string}\rightarrow z_p,& \mathcal{D}_p&:z_p\rightarrow \text{plan string}.
\end{align*}
At inference time, decoding $z$ yields an executable pair $(\widehat{Q},\widehat{P})$; the plan string is turned into database-specific instructions (such as an optimizer hint string~\cite{url-pg_hints}) and combined with $\widehat{Q}$ for execution. 

\paragraph{Plan representation and VAE.} Our plan representation is identical to prior work using Bayesian optimization for query planning~\cite{bayesqo}. At a high level, we encode plans as strings of integers which are then interpreted as operator selections and join orders. Like SELFIES~\cite{SELFIES}, every string represents a valid plan, and every plan is represented by at least one string, although (unlike SELFIES) some plans are represented by more than one string. This means that any sequence decoded from $\mathcal{E}_p$ can be interpreted as a valid plan.

\paragraph{Query VAE.}
Unlike query plans, SQL lacks a SELFIES-style~\cite{SELFIES} closed vocabulary where any token sequence is valid. To sidestep data scarcity ($\sim$20k queries) and grammar complexity, we do not train a full encoder–decoder VAE for queries. Instead, we (i) use a text embedding model as the query encoder $\mathcal{E}_{q}$ (OpenAI \texttt{text-embedding-3-large}), then (ii) fine-tune a small LLM (\texttt{Qwen-2.5-0.5B}) as a compact query decoder $\mathcal{D}_{q}$ to map the first 256 embedding dimensions back to SQL queries. This is similar to prompt tuning \cite{prompt_tuning}, where the embedding is passed to the LLM as a soft-prompt, and the LLM is trained to decode the embedding back to the original query string. We use grammar-constrained decoding during inference to guarantee syntactic correctness of the reconstructed query. This yields $\sim$67\% string reconstruction accuracy for queries and $\sim$99\% for plans. 

\paragraph{A Simple SQL Subset} To generate simple queries that work across a wide variety of databases, we focus on conjunctive queries~\cite{yannakakis_plus}. Essentially, queries are restricted to: (1) joining together relations with equijoins (e.g., between foreign and primary keys), and (2) conjunctive predicates on table columns. Each query performs an ungrouped \texttt{count(*)} aggregation. Our formulation thus excludes SQL features like subqueries, window functions, recursion, etc. While simple, conjunctive queries are still difficult for modern query optimizers~\cite{howgood}.

\paragraph{Grammar-constrained decoding for SQL.}
To make sure the query decoder $\mathcal{D}_{q}$ always decodes to a valid query, we compiled a concise EBNF for our SQL subset (by enumerating the powerset of joinable tables and filterable columns) and enforced it during $\mathcal{D}_{q}$ decoding via grammar constrained generation using vLLM. This addresses the main challenges associated with unconstrained decoding: (i) syntactic invalidity and (ii) semantically impossible references (e.g., undefined aliases). The grammar constraint only applies to $\mathcal{D}_{q}$, since  $\mathcal{D}_{p}$ is designed to have a set of closed vocabulary where any decoded sequence is always valid.

\section{Preliminary results}

To test the feasibility of our approach, we generated a workload against the IMDb dataset (chosen to facilitate comparison with prior work using the same dataset). To match the original join order benchmark~\cite{howgood}, we generated 122 query plans. These plans were collected by selecting the best candidates (in terms of relative and absolute headroom) from a weeklong BO run. Queries were executed on DuckDB~\cite{duckdb} on a node equipped with an AMD Ryzen 5 3600 and 64GB of RAM. The entire dataset was cached in memory.

\begin{figure}
    \centering
    \includegraphics[width=0.7\linewidth]{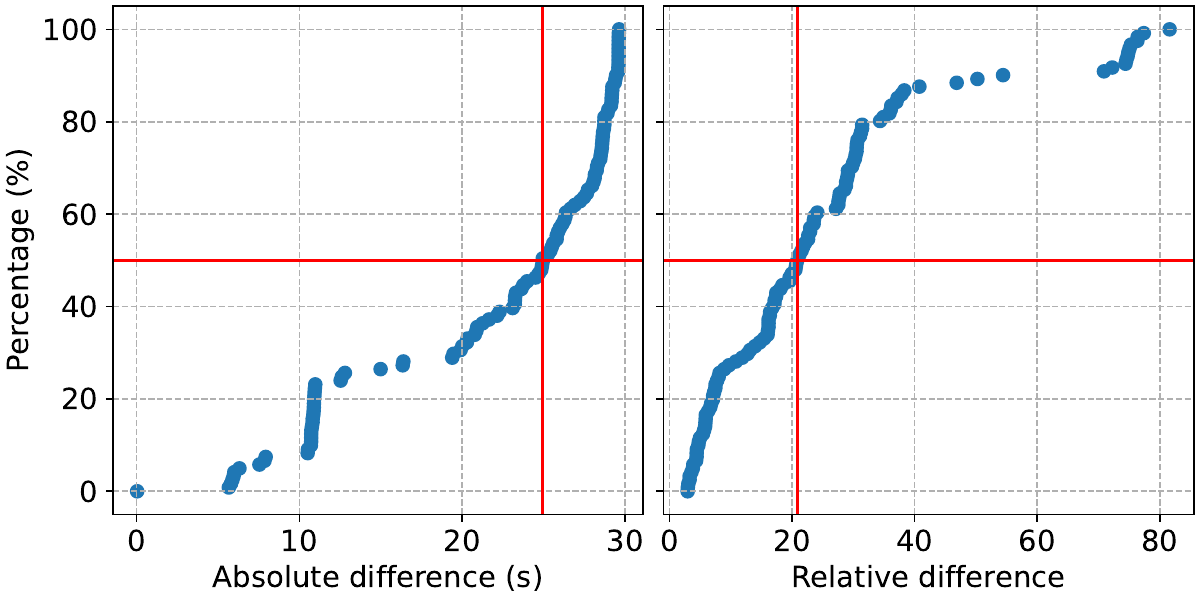}
    \caption{CDFs of absolute and relative differences between baseline system (DuckDB) and witness plan for our generated benchmark. Median values at red cross.}
    \label{fig:cdfs}
\end{figure}

We plot the CDFs of the absolute and relative headrooms in Figure~\ref{fig:cdfs}. The median query in our generated workload had an absolute headroom of 25 seconds, meaning that the witness plan was 25 seconds faster than the default plan chosen by DuckDB. In terms of relative headroom, the median query in our generated workload had a witness plan that was 20x faster than the default plan chosen by DuckDB. The largest differences found approach 30 seconds in absolute terms, and nearly 80x in relative terms. The least difficult query in our workload has less than a second of absolute improvement and only a 1.5x relative improvement (potentially attributable to noise). We hope that longer BO runs in the future will continue to shift both distributions right; the comparative ``easiness'' of the weakest generated queries is potentially attributable to a lack of search time.

\begin{figure}
    \centering
    \includegraphics[width=0.5\linewidth]{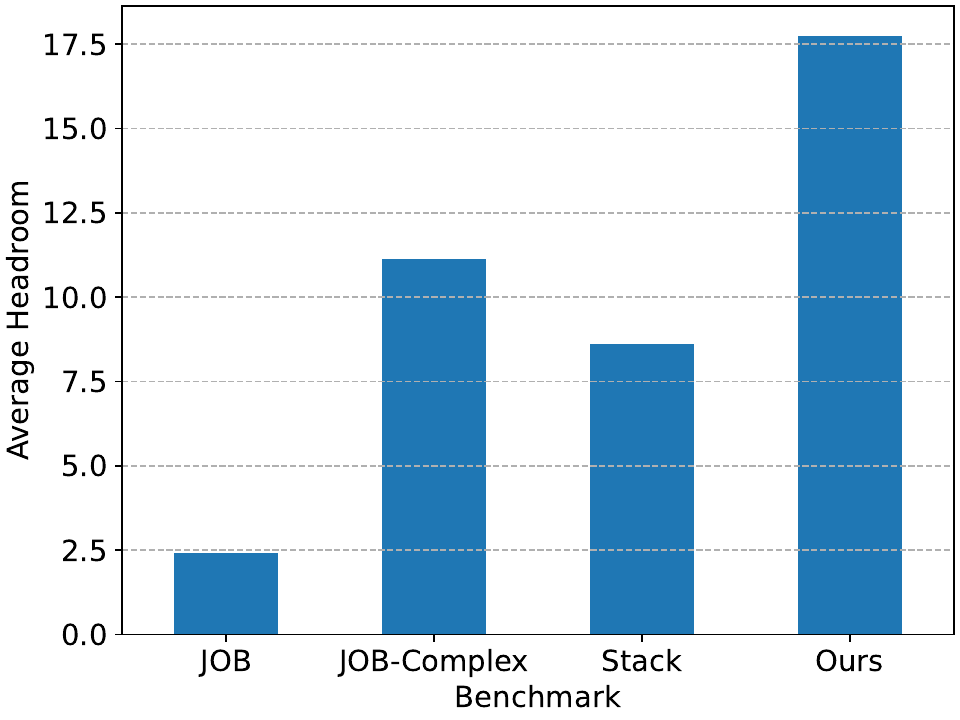}
    \caption{Comparison of geometric mean relative headroom for four benchmarks. Values for JOB and JOB-Complex are from~\cite{job_complex}. Values for Stack are approximated from~\cite{bao}. Values for ``Ours'' are computed with respect to the witness plan.}
    \label{fig:compare}
\end{figure}

To compare the quality of our generated benchmark against prior work, Figure~\ref{fig:compare} shows the mean (geometric) relative headroom of different benchmark. Since computing the headroom of a benchmark is expensive (i.e., one must execute millions of query plans per query), we rely on results reported in prior work, and thus compare only the mean relative headroom (as absolute headroom is not consistently reported).

\section{Conclusion and Future Work}

In this extended abstract, we presented a fully automated approach for creating challenging database benchmarks by viewing benchmark creation as an optimization problem and using Bayesian optimization in a joint latent space over both SQL queries and execution plans. Our combination of VAE decoders enables joint embeddings for queries and plans, while grammar-constrained decoding ensures syntactic validity of generated SQL within a simple, widely portable subset (conjunctive queries with equijoins and predicates). Together, these choices let the system efficiently find adversarial (query, plan) pairs with substantial optimization headroom.

Preliminary experiments on IMDb with DuckDB show promising evidence: from a week-long BO run we selected 122 candidates, and the resulting workload exhibits a median absolute headroom of ~25 s and a median relative headroom of ~20× between the default plan and the discovered witness plan (with the largest gaps approaching ~30 s and ~80×, respectively). These results suggest our automated process can find queries that are a lot harder than those in commonly used benchmarks. 

For future work, with a much larger set of hard (query, plan) pairs from extended runs, we could fine-tune a schema-conditioned LLM to directly generate pairs in bulk, skipping Bayesian optimization for most cases. The model would learn from (SQL, plan, headroom) triples, use grammar-constrained decoding to preserve validity, and be paired with a fast verifier (parse + quick execution/cost check) to filter outputs. A simple reward model or DPO-style objective that favors higher headroom can bias generation toward adversarial regions without explicit search. In practice, we could batch-generate candidates, keep the top-k under the verifier, and fall back to optimization only when confidence is low, therefore cutting wall-clock time while maintaining quality through periodic active-learning sessions.

Overall, our method reduces manual effort, directly targets “difficulty” via headroom, and yields concrete witness plans that can help with debugging and system improvement, while keeping queries broadly executable across systems.

\newpage
\begin{ack}
N. Maus was supported by the National Science Foundation Graduate Research Fellowship; 
J. R. Gardner was supported by NSF awards IIS-2145644 and DBI-2400135;

\end{ack}

\bibliographystyle{plainnat}
\bibliography{references,ryan-cites-long}





\end{document}